# SIMULATION OF CRAB WAIST COLLISIONS IN DAΦNE WITH KLOE-2 INTERACTION REGION

Mikhail Zobov, Alexander Valishev, Dmitry Shatilov, Catia Milardi, Antonio De Santis, Alessandro Drago, Alessandro Gallo

*Abstract*— After the successful completion of the SIDDHARTA experiment run with crab waist collisions, the electron-positron collider DAΦNE has started routine operations for the KLOE-2 detector. Similarly to the SIDDHARTA configuration, the new interaction region exploits the crab waist collision scheme, but features certain complications including the experimental detector solenoid, compensating anti-solenoids, and tilted quadrupole magnets, that lead to significant coupling of the horizontal and vertical betatron motion. It is not immediately obvious if the crab waist scheme would be as efficient in the strongly coupled case as it was in the uncoupled configuration. We have performed simulations of beam-beam interactions in the collider taking into account the real machine nonlinear lattice. In particular, we have evaluated the effect of crab waist sextupoles and beam-beam interactions on the collider dynamical aperture and energy acceptance. A new betatron tune working point has been proposed for the DAΦNE electron ring, and its implementation resulted in more than 20% background reduction and injection efficiency improvement. Exploiting this working point has allowed reaching the best present luminosity of $2.0\times10^{32}$ cm$^{-2}$s$^{-1}$. The numerical simulations have shown that for the given bunch currents in collision, the powering of the crab waist sextupoles should decrease the beam core blow up by a factor of 2 indicating that even higher luminosity can be achieved in DAΦNE thus encouraging further collider optimization.

## I. INTRODUCTION

DAΦNE is an accelerator complex the main element of which is a double ring electron-positron collider operating at the center-of-mass energy of the Φ-resonance (1.02 GeV) [1]. The implementation of the so-called Crab Waist collision scheme (CW) has led to the achievement of the record peak luminosity $L=4.5\times10^{32}$ cm$^{-2}$s$^{-1}$ while working for the SIDDHARTA experiment [2]. This success has led to a decision to reinstall the upgraded KLOE detector, KLOE-2, exploiting the advantages of the CW scheme. For this purpose, the KLOE Interaction Region (IR) was carefully redesigned [3], and in 2013 the KLOE detector has been upgraded with new layers in the inner part of the apparatus [4]. The new KLOE-2 IR is much more complex as compared to the SIDDHARTA IR because the collisions take place inside the detector solenoid, IR has rotated quadrupoles and additional compensator solenoids.

At present, the luminosity of $2.0\times10^{32}$ cm$^{-2}$s$^{-1}$ has been achieved to be compared with the maximum of $1.5\times10^{32}$ obtained in the previous KLOE run [5]. Table I summarizes the most relevant machine and beam parameters. It is worth noting that the performance is mostly limited by collective effects such as electron cloud, microwave instability, feedback system noise, etc. [6]. An extensive campaign of experimental and simulation studies is in progress in order to fully exploit the collider potential and push the luminosity to higher values.

Numerical simulations of beam-beam effects with the weak-strong particle tracking code Lifetrac proved to be very efficient for the detailed understanding of the CW collision scheme, and for the collider performance optimization [7,8]. The goal of the present work was to expand modeling to the case of strong betatron coupling by implementing the complete nonlinear collider lattice with IR inside of a strong detector solenoid in weak-strong beam-beam simulation. We then apply computer simulation to demonstrate the effectiveness of crab waist collision scheme and to guide parameter optimization, such as a better working point choice, crab sextupole strength optimization, correction of the phase advance between the sextupoles and the interaction region. Where possible, we apply the simulation results to collider operation.

In addition to the importance for the DAΦNE program, the studies have general value since they represent a unique opportunity to benchmark the numerical tools against the experimental data and enable further applications towards future colliders, such as the High Luminosity upgrade of the Large Hadron Collider (HL-LHC) [9] and the Future Circular Collider (FCC) [10]. The conclusion about the effectiveness of the crab waist collision scheme is essential for future circular electron-positron colliders.

This work was supported by the U.S. Department of Energy via the US-LARP program, and by the European Union FP7 HiLumi LHC, Grant Agreement 284404.

Alexander Valishev is with Fermi National Accelerator Laboratory, Batavia, IL 60510 USA (e-mail: valishev@fnal.gov).

Mikhail Zobov, Catia Milardi, Antonio De Santis, Alessandro Drago, Alessandro Gallo are with Laboratori Nazionali di Frascati of Istituto Nazionale di Fisica Nucleare, 00044 Frascati (Roma) Italy.

Dmitry Shatilov is with Budker Institute of Nuclear Physics, 630090 Novosibirsk, Russia.

## II. SIMULATION TOOLS

Lifetrac is a weak-strong particle tracking code that was originally developed for the simulation of equilibrium density distributions in lepton colliders [7], and was later expanded to enable non-equilibrium simulations for hadron machines [11,12]. The fully symplectic 6D treatment of beam-beam interaction allows to perform simulations with very large crossing angle, and crabbing of either weak and strong bunches. The capabilities of the code in the equilibrium distribution case include: (i) the ability to calculate 3-D density of the weak beam; (ii) evaluate the specific luminosity and beam lifetime; (iii) calculate the area of stable particle motion (dynamical aperture, DA); (iv) perform Frequency Map Analysis (FMA) [8].

TABLE I
DAΦNE MACHINE AND BEAM PARAMETERS DURING 2015 OPERATION WITH KLOE-2 DETECTOR (APRIL 2015)

| Parameter | Value |
| --- | --- |
| Beam energy | 510 MeV |
| Circumference | 97 m |
| Luminosity | $2.0\times10^{32}$ cm$^{-2}$s$^{-1}$ |
| Number of bunches | 95 |
| Bunch spacing | 2.7 ns |
| Full horizontal crossing angle | 50 mrad |
| Number of electrons and positrons per bunch | $2.05\times10^{10}$ |
| Beam emittance, r.m.s. (x,y) | 0.28, 0.0021 $\times10^{-6}$ m |
| Momentum spread, r.m.s. | $5.5\times10^{-4}$ |
| Bunch length, r.m.s. (e$^-$, e$^+$) $\sigma_z$ | 1.55, 1.6 cm |
| Electron betatron tunes ($\nu_x,\nu_y$) | 0.130, 0.170 |
| Positron betatron tunes ($\nu_x,\nu_y$) | 0.098, 0.130 |
| Damping decrements (x,y,z) | (1.1, 1.1, 2.2) $\times10^{-5}$ |
| Beta-function at IP (x,y) | 25, 0.84 cm |
| Beam-beam parameter, e$^+$ (x,y) | 0.011, 0.04 |

A recent improvement of the code relevant for the present study was the implementation of detailed machine lattice model via element-by-element tracking in thin lens approximation. The optics data for both the weak and the strong beam are imported from MAD-X [13] model files where the element slicing is performed using the methods available in MAD-X. For the purpose of this work the slicing was done with the so-called Teapot algorithm [14]. The thin lens tracking is implemented following [15], and makes use of the paraxial approximation for the multipole elements and properly treats non-paraxial effects in the drifts. This method proved to be accurate for the LHC tracking studies and is appropriate for the DAΦNE case. Such approach enables the proper treatment of element misalignments and related orbit distortions, chromatic aberrations, lattice nonlinearities, betatron and synchro-betatron coupling.

## III. SIMULATION RESULTS

### A. Dynamical Aperture

The achievement of maximum DA is essential for the improvement of injection efficiency, beam lifetime, and background conditions for the experiment. Figs. 1,2 present the results of FMA and DA simulations for the e$^-$ beam in the absence of beam-beam interaction. The colour on FMA plots represents the tune jitter in logarithmic scale from red ($10^{-3}$) to blue ($10^{-7}$). The stability boundary established in long-term ($2\times10^5$ turns) tracking agrees well with the FMA data. Note that the DA substantially decreases for off-momentum particles as a consequence of uncorrected chromatic aberrations near the half-integer or the integer resonance. For DAΦNE betatron tunes farther from the integer should be favored unless the sextupoles are optimized to correct higher-order lattice aberrations [16].

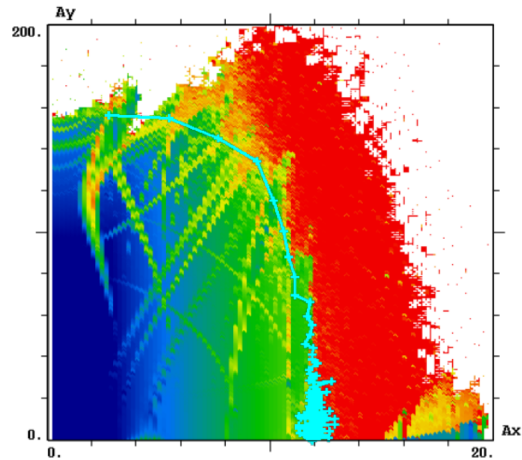

Fig. 1. FMA (colour chart) and $2\times10^5$ turns DA (cyan line) for the e$^-$ beam without beam-beam interaction. Normalized synchrotron amplitude $A_s$=1. Betatron tunes ($\nu_x$=0.098, $\nu_y$=0.164). Horizontal and vertical axes are labelled in units of the respective beam size.

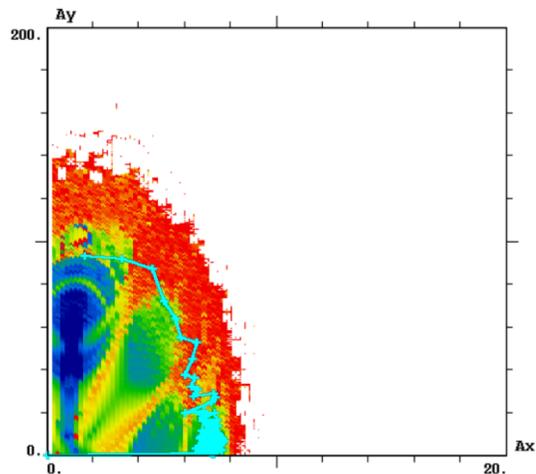

Fig. 2. Same as Fig. 1 with normalized synchrotron amplitude $A_s$=9.

### B. Working Point Optimization

We performed betatron tune scans for both the e$^+$ and e$^-$ rings to determine the optimal running condition in order to achieve the maximum specific luminosity and dynamical aperture (beam lifetime). In Fig. 3, the inverse beam size is plotted. The simulations established that the working point for the positron ring was near optimal, while for the electron ring the DA was insufficient. The proposed new working point provides a significantly better DA at the same or marginally better specific luminosity (Figs. 4-6). Fig. 4 suggests that horizontal DA in the new working point improves by about 2-3σ, and from Figs. 5(a),6(a) one sees that the vertical tail growth is suppressed. These changes should result in a better injection efficiency and lower particle losses. Figs. 5(c) and 6(c) demonstrate that the core size in the old and the new working point is essentially the same, and one should expect minimal impact on the specific luminosity.

The numerically suggested working point was implemented in the DAΦNE electron ring. A substantial background reduction was observed immediately after the new optics application and multibunch feedback systems tuning. In particular, in the left panel of the Fig. 7, the comparison between the machine normalized backgrounds hitting the KLOE-2 calorimeter in the region around the exit of the electron beam is shown as a function of the instantaneous luminosity. The reduction of the normalized background generated by the electron beam ranges between 30% and 20%. In the right panel of Fig. 7 the effect of the electron beam optics variation is shown for the KLOE-2 normalized trigger rate. Note that the DAΦNE background is dominated by the Touchek effect depending on the bunch volume and inversely proportional to the bunch intensity. Since the transverse beam sizes increase due to the beam-beam interaction and grow above the transverse mode coupling instability threshold, and the bunch lengthens due to the potential well effect and the microwave instability, in Fig. 7 we observe some reduction of the normalized background and trigger level for the higher luminosity. The observed improvement of the trigger rate has

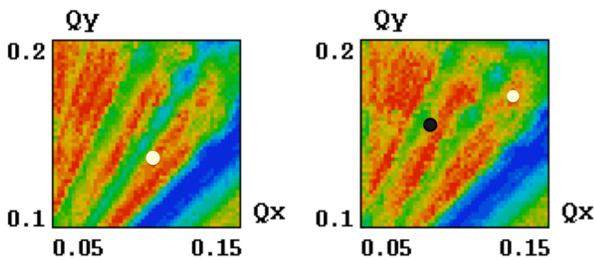

Fig. 3. Factor $1/\sqrt{\varepsilon_x\varepsilon_y}$ in colour (red – large, blue – small, the range corresponds to a ten-fold change) as a function of betatron tune for positrons (left), and electrons (right). White point shows the optimal working point; black dot is the original e⁻ tune.

two different positive effects:

a) The data throughput is decreased because the contribution of the machine background hitting the KLOE-2 sub-detectors to the event size is smaller and because the rate

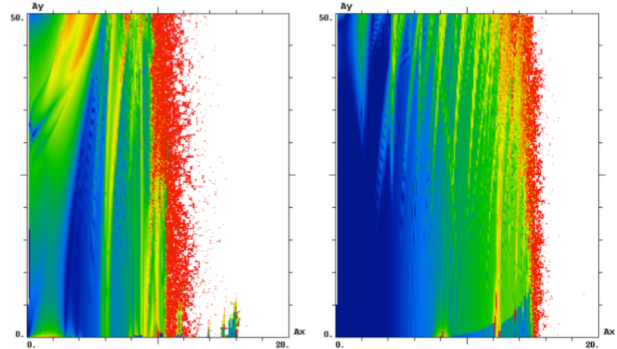

Fig. 4. FMA for electron beam particle with energy offset of 2 σ, beam-beam interaction off. Working point $v_x$=0.088, $v_y$=0.152 (left), $v_x$=0.130, $v_y$=0.170 (right).

of events acquired is smaller.

b) The dead time induced by the activation of the KLOE-2 trigger is smaller allowing for a more efficient data taking.

The reduction of the machine background rate is of primary interest for the KLOE-2 experiment, immediately after the increase of delivered luminosity, because it has a very high impact on all subsystems involved in the data acquisition: DAQ boards, KLOE-2 networks, data buffering, mid-term (disk) and long-term (tape) consumption, etc. For these reasons the result achieved with the new optics represent exactly the kind of machine development that is needed for the KLOE-2 data taking.

In addition to the reduced background, after few days of collider fine-tuning, the best KLOE-2 experiment peak luminosity of $2.0\times10^{32}$ cm$^{-2}$s$^{-1}$ and daily integrated luminosity of 11 pb$^{-1}$ have been achieved.

C. *Effect of Crab Waist*

Simulations with the detailed machine optics clearly demonstrate the benefits of the CW scheme in the KLOE-2 configuration of DAΦNE. In both working points, switching the CW sextupoles off results in a significant blow-up of the beam core, and about a two-fold decrease of the specific

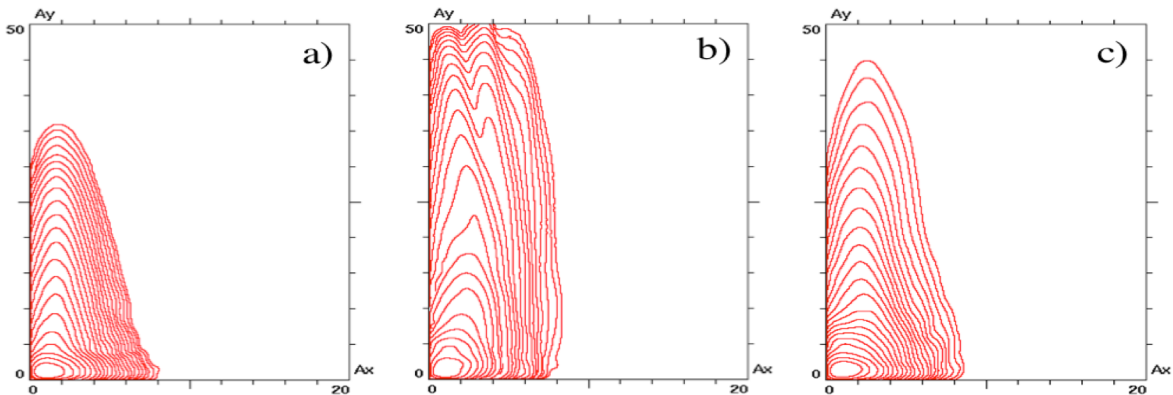

Fig. 5. Equilibrium density contour plots in the betatron amplitude space for electron beam in the working point vx=0.088, vy=0.152. a) – beam-beam off; b) – with beam-beam, CW off; c) – with beam-beam, CW on.

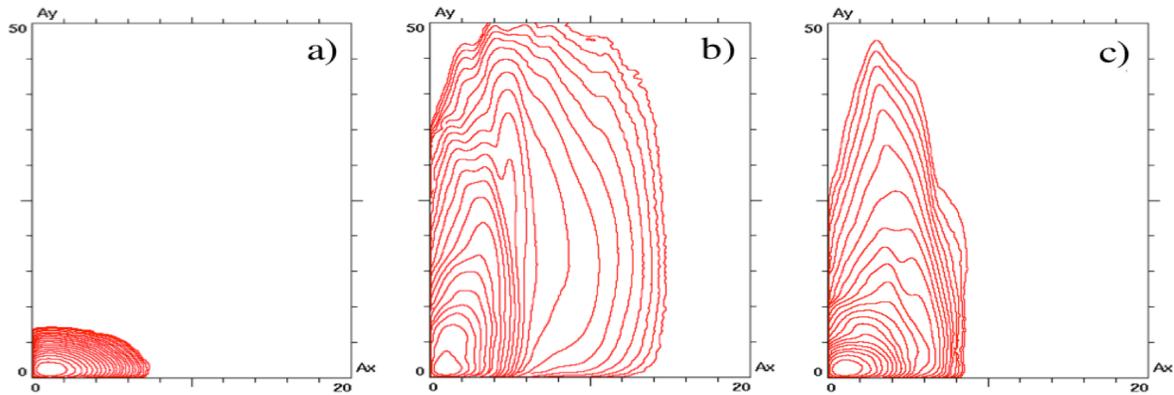

Fig. 6. Equilibrium density contour plots in the betatron amplitude space for electron beam in the working point $v_x=0.130$, $v_y=0.170$. Left – beam-beam off; middle – with beam-beam, CW off; right – with beam-beam, CW on.

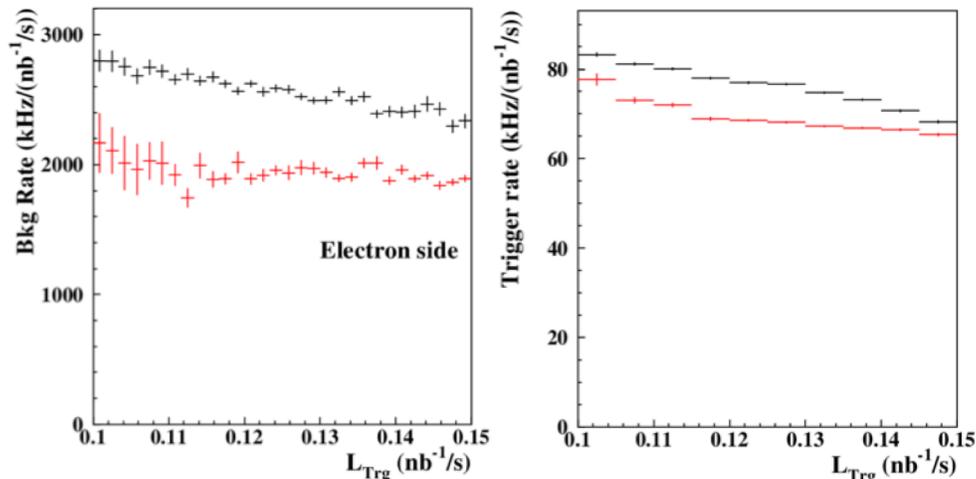

Fig. 7. Comparison between the machine normalized background hitting the KLOE-2 calorimeter in the region around the electron beam exit as a function of instantaneous luminosity (left) and KLOE-2 normalized trigger level (right) for the previous (black dots) and the current (red squares) electron machine working point. KLOE-2 trigger provides the instantaneous luminosity measurement. The reduction of normalized background rate as a function of luminosity for the previous DAΦNE optics is mainly due to the dynamical effect induced by the high current (~1A) needed to reach high luminosity. The reduction in the KLOE-2 normalized trigger rate observed in the right panel has a similar behavior as the background rate.

luminosity (Figs. 5(b)-5(c) and 6(b)-6(c)).

A comparison of the simulations with simplified linear lattice and the detailed lattice yields that for good working points the nonlinearities and the experimental detector solenoid do not impact significantly the collider performance resulting in less than 15% luminosity reduction.

The beneficial effect of the crab sextupoles is clearly seen in operations [17]: when increasing the strength of the sextupoles the vertical beam size decreases, and, respectively, the measured luminosity increases. However, the potential of the crab waist collision scheme has not been fully exploited yet due to several beam dynamics issues such as the electron cloud effect, single bunch instabilities etc. The limiting factors and future plans to overcome them are discussed in detail in [17].

## IV. CONCLUSIONS

The weak-strong simulations of beam-beam effects in DAΦNE taking full account of the coupled machine optics with KLOE-2 interaction region clearly demonstrate the advantage of the crab waist collision scheme.

Simulations established that the configuration of the $e^+$ ring is near optimal, while the dynamical aperture of the $e^-$ ring could be improved by a change of the working point. In full agreement with the model prediction, the new $e^-$ working point resulted in a substantial improvement of the injection efficiency, beam lifetime and background conditions, with a moderate increase of the specific luminosity. In particular, the fine tuning of the injection system parameters at each electron beam injection (after re-switch of the system from the positron injection mode to the electron beam injection) is no longer necessary due to the improvement in dynamic aperture and momentum acceptance at the new working point. A further 30% increase of the luminosity is possible through the proper alignment of CW sextupoles, correction of the local coupling in IR, and optimization of CW sextupole strength.

In addition to the importance for the DAΦNE program, our studies have general value. On the one hand, the numerical simulations have clearly demonstrated that the crab waist scheme works well even if the beam collisions take place inside the experimental detector with a strong solenoidal field. This is important for promoting studies of future electron-positron colliders based on the crab waist collisions scheme, FCC-ee crab waist option, for example [18]. On the other hand, the benchmarking of the numerical simulations against the DAΦNE operating experience makes us more confident in our numerical results, conclusions and predictions while applying our numerical tools for beam-beam interaction studies for the High Luminosity upgrade of the LHC (HL-LHC).

ACKNOWLEDGMENT

We are indebted to the DAΦNE technical staff for help in the machine development. The authors D. Shatilov and A. Valishev would like to thank the DAΦNE team for hospitality during their visits.